# Stakes are higher, risk is lower: Citation distributions are more equal in high quality journals[1]

Andrey Lovakov [*] and Vladimir Pislyakov[**]

[*]*lovakov@hse.ru*
Center for Institutional Studies, National Research University Higher School of Economics, Myasnitskaya 20, Moscow, 101000 (Russia)

[**] *pislyakov@hse.ru*
Library, National Research University Higher School of Economics, Myasnitskaya 20, Moscow, 101000 (Russia)

## Introduction

Several recent bibliometrics studies have reignited the well-known debates initiated more than twenty years ago by vivid works of Per Seglen (1992; 1994; 1997). The question is whether impact factor may represent not only the citedness of a journal as a whole, but also give some estimate of individual papers' quality published in it (different views: Larivière et al., 2016; Zhang, Rousseau & Sivertsen, 2017; Waltman & Traag, 2017; Pudovkin, 2018). This is an important and profound theme of interrelation between a part and a whole, their mutual dependency and the limits of this dependency.

To explore this research question, we analyze correlation between the average (in our case, journal impact factor, IF) and the amplitude of oscillations/deviations around this average (citations received by individual papers in the journal). This is, so to say, "indicators of the second order", we measure the digression of the citations received by individual papers from the journal's average. To evaluate this digression a Gini coefficient is used. This index ranges from 0 to 1 where 0 corresponds to perfect equality (every paper in a journal receives the same number of citations) and 1 corresponds to perfect inequality (all citations are received only by one single paper).

Several previous works already addressed the question of relationship between IF and Gini coefficient. Some studies explored particular disciplines, e. g. dermatology (Stegmann & Grohmann, 2001) or cardiovascular research (Nuti et al., 2015). Others introduced mixed indices combining average number of citations and dispersion measures (Lando & Bertoli-Barsotti, 2017; Cockriel & McDonald, 2018).

We try to put it in a different perspective. In addition to answering the basic question 'how good IF can represent individual journal articles', we reflect on possible application of our findings by scientists who plan and develop their publication careers.

Our previous study (Lovakov & Pislyakov, 2017) was focused on psychology journals and used division of journals by their "functional types" (methodology journals, sub-discipline journals etc.). There we have received controversial results, with no uniform correlation between Gini and IF.

[1] The paper is written within the framework of the Basic Research Program of the National Research University Higher School of Economics and was supported by the Russian Academic Excellence Project '5–100'.

In this paper we again use psychology journals as a starting point. Psychology is a discipline standing at the crossroads of hard and social sciences. Some of psychology journals are attributed to Science Citation Index in the Web of Science database while others to Social Sciences Citation Index, and some to both. But now, first, we move away from functional typology of journals within the same science domain and explore them as a consolidated set of sources. Next, we pay attention to the most frequent question which was evoked by the reviewers/readers/listeners of (Lovakov & Pislyakov, 2017). Do the regularities found for psychology are also valid in hard sciences and in social sciences?

To answer, we shift the focus of our study towards more "hard" and more "social" counterparts of psychology. The natural sciences extreme of psychology may be represented by neuroscience. The shift to social sciences may lead us to sociology.

Why we choose these disciplines as "close neighbors". Psychology studies a mind and behavior, while neurosciences study a nervous system and in the first place the brain which is a biological basis for mind and behavior. A journal *Psychology and Neuroscience*, published by American Psychological Association, is an additional proof that these fields are close to each other.

As for shift toward social sciences, we may note that sociology studies society as a whole, and social psychology (social part of the field) studies how social groups and society affect the thoughts and behavior of individual persons. There is even a special category *Psychology, Social* in JCR ontology.

**Data and Methods**

Data on journals from the three fields of science (Psychology, Neurosciences, and Sociology) indexed in Science Citation Index Expanded (SCIE) and Social Sciences Citation Index (SSCI) (WoS—Web of Science platform, Clarivate Analytics company) were used. Psychology is considered as a combination of all journals from 10 WoS/JCR categories: Psychology; Psychology, Applied; Psychology, Biological; Psychology, Clinical; Psychology, Developmental; Psychology, Educational; Psychology, Experimental; Psychology, Mathematical; Psychology, Multidisciplinary; Psychology, Social. Neurosciences field includes all titles from WoS/JCR category Neurosciences. Sociology consists of the journals from WoS/JCR category Sociology. Forty journals were randomly selected from each field (10 titles from each quartile identified by average journal impact factor percentile, definition is given below).

Two sets of data were collected for each journal. First, four journal indicators were extracted from 2015 edition of Journal Citation Reports (JCR, Clarivate Analytics): two-year and five-year journal impact factors, journal impact factor without journal self cites, average journal impact factor percentile (average IF percentile if a journal is included to several discipline categories). Second, data about citations in year 2015 of all papers published by selected journals in 2010–2014 were extracted from WoS Citation Report (only 'article' and 'review' document types were taken into analysis). Then five separate Gini indexes were calculated for each journal. Each of them was based on citations received in 2015 to by papers published in each single year from 2010 to 2014. From these five values two average Gini indexes were aggregated using their arithmetic mean. One includes two Ginis of 2014 and 2013 publications (2-year average Gini), another takes all five Ginis for papers published from 2014 to 2010 (5-year average Gini). This approach allows to more consistently compare these averages with two-year and five-year 2015 impact factors as the same papers/citations are assessed by them: 2015 citations of articles/reviews published from 2014 to 2013 and from 2014 to 2010 appear in the formulae of 2-year and 5-year impact factors-2015 respectively. As an additional variable 2-year impact factors without journal self-citations were taken from JCR.

At last, for comparison with Gini an average IF percentile was collected. This indicator is also calculated by JCR. An IF percentile is a share of journals from the same discipline which IFs are lower than the IF of a given journal (so, the better is journal, the higher is percentile). An average IF percentile is an average of percentiles' values if the journal is attributed to more than one discipline and its IF percentiles vary among disciplines.

For each journal we also calculated the share of papers which were published in 2010–2014 and have not received any citations till the end of 2015, so called "uncitedness factor" (van Leeuwen & Moed, 2005; Egghe, 2008).

For the data preparation, analysis and visualization we used R, a programming environment for statistical computing (R Core Team, 2017).

**Results and Discussion**

*Descriptive statistics*

Table 1. Gini index and journal indicators based on impact factor.

|  | Psychology | | Neuroscience | | Sociology | |
|---|---|---|---|---|---|---|
|  | min | max | min | max | min | max |
| Two-year average Gini index | 0.44 | 0.84 | 0.25 | 0.92 | 0.45 | 0.94 |
| Five-year average Gini index | 0.35 | 0.87 | 0.10 | 0.89 | 0.44 | 0.92 |
| Two-year journal impact factor | 0.18 | 6.71 | 0.14 | 11.21 | 0.04 | 2.27 |
| Five-year journal impact factor | 0.19 | 10.77 | 0.14 | 10.80 | 0.02 | 5.62 |
| Journal impact factor without journal self cites | 0.04 | 6.71 | 0.06 | 10.86 | 0.00 | 1.99 |
| Average journal impact factor percentile | 1.57 | 95.74 | 0.59 | 96.36 | 1.06 | 91.20 |

*Gini index and percentage of uncited papers*
Uncitedness coefficient (factor) was analyzed in a plenty of studies, a concise bibliography was collected by Haustein (2012).

We found that the Spearman rank correlation coefficients of Gini index and percentage of uncited papers are positive and high for psychological, neurosciences, and sociological journals (0.82, 0.71, and 0.94, respectively). The higher percentage of uncited papers, the higher inequality in citations distribution, Figure 1 shows their relationship. For neurosciences journals this correlation is weaker, which means that there are journals with quite a lot of uncited papers, but other articles/reviews in them are cited rather equally.

Figure 1. Relationship between 5-year average Gini index and percentage of uncited papers.

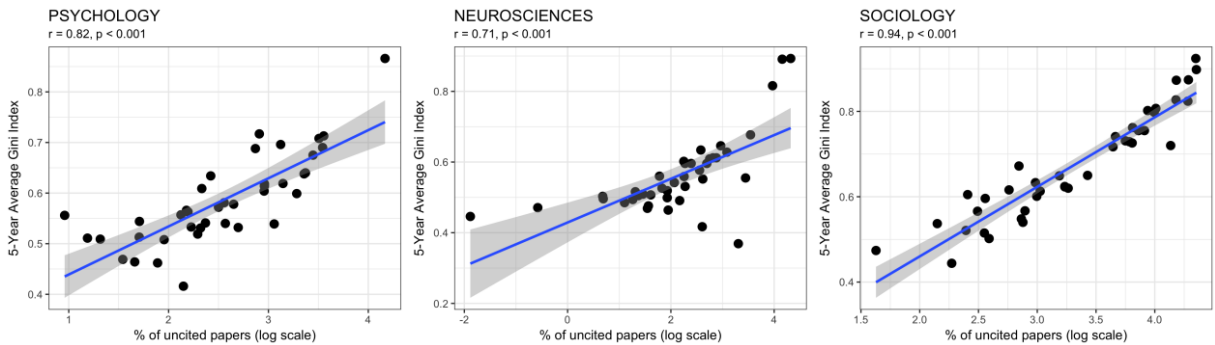

*Gini index and journal impact factor*
To find the relationship between inequality of a citation distribution across a journal and its "quality" measured by IF, the Pearson correlation coefficients for Gini index and four journal indicators based on journal impact factor were calculated.

Figure 2. Average Gini index and four journal indicators based on impact factor
of 40 psychological journals.

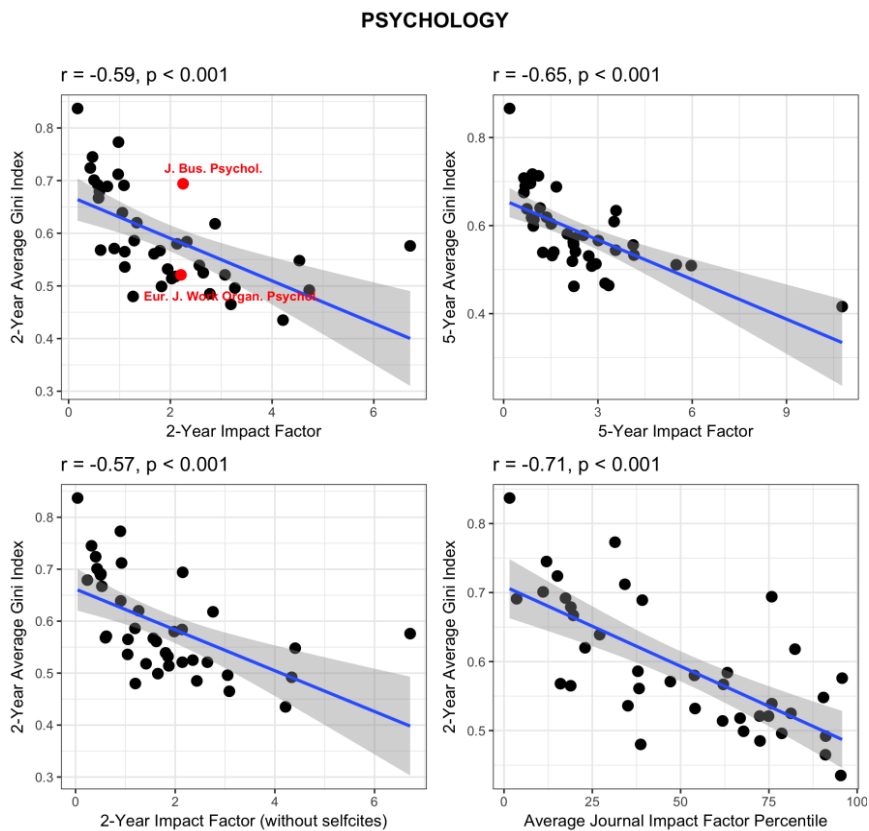

In all fields a negative correlation between Gini and journal indicators is observed. Pearson correlation coefficients vary from −0.57 to −0.71 for psychological journals, from −0.61 to −0.82 for neurosciences journals, and from −0.78 to −0.79 for sociological journals. Figures 2–4 show relationship between Gini index and four journal indices based on impact factor. Dots are journals, regression lines are shown with 95% confidence intervals for predictions from a linear model.

Figure 3. Average Gini index and four journal indicators based on impact factor of 40 neurosciences journals.

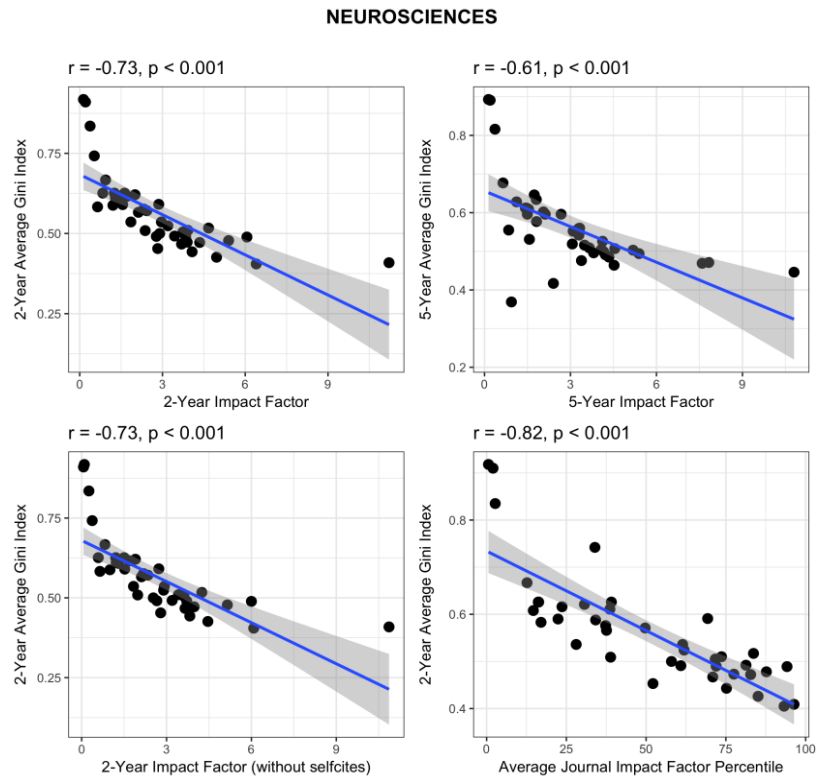

Figure 4. Average Gini index and four journal indicators based on impact factor of 40 sociological journals.

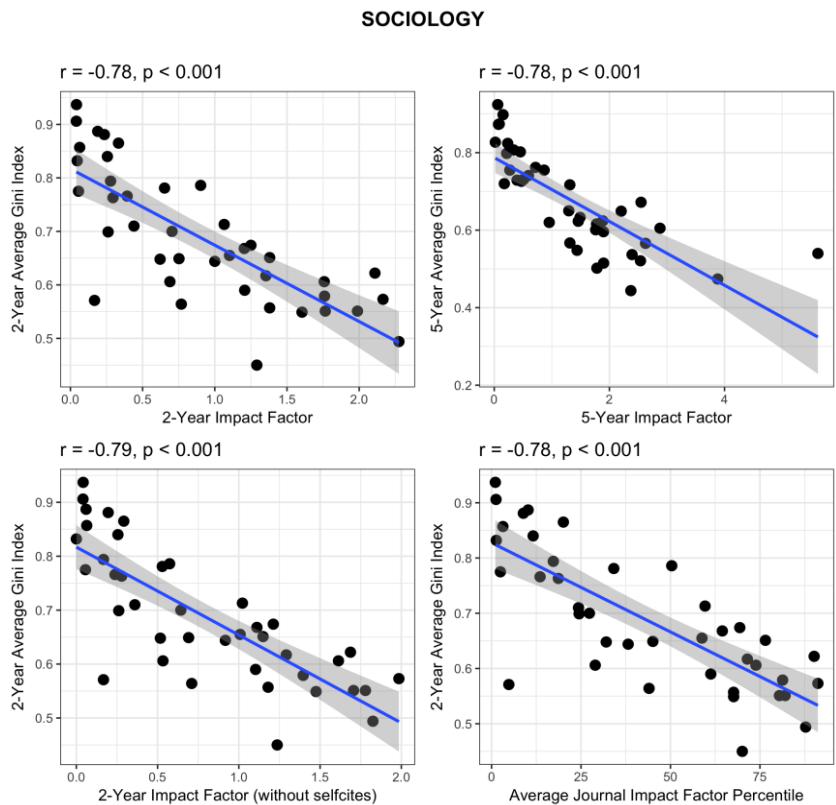

These results demonstrate the robust relationship between journal impact factor and Gini index. The higher the impact factor of the journal, the more equally the citations are distributed among its papers.

Several hypotheses explaining the observed regularities may be formulated. First of all (let us call it "editorial hypothesis"), it may be suggested that the editors/referees of the most prestigious journals select only those manuscripts which will be actively read and cited. These journals have a high "rejection rate", they hold a great number of submitted manuscripts and so may choose those which will have an impact on scholarly audience.

Next, the "readership hypothesis". A reader (whether scientist or a person interested in the psychology/neurosciences/sociology) knows that nowadays it is impossible to track all knowledge flow in his science domain. He has to choose, and the rational choice must take into account the experience of the other actors. "First of all, investigate those sources which were thoroughly investigated by other experts". In practice this means "read/look through high impact journals from cover to cover". Single papers published in less prominent sources may receive a proportionate attention only if they were noticed by opinion leaders, incited a discussion, the results were indexed by academic search platforms, bibliometric databases etc. But these are "particular stories", not the general line. So to say, it is in high impact journals where the "journal identity" matters, in low percentiles there are only individual articles' cases.

At last, another explanation may be provided by a "journal's perspective". Journals with high IFs receive a greater number of citations and their individual papers are less dependent on random outliers and unexpected popularity. Their overall indicators tell more about *each paper* than in the case of titles with low average citedness per document. Nuti et al. (2015) put it in a more decisive manner: "Our findings reassure readers that IFs are generally representative of the quality and contribution of the entire spectrum of a journal's papers and not just a select few" (here with not well-founded extension to poorly cited journals).

Anyhow, the latter hypothesis as well as our findings dispute or at least mitigate now widespread critics of the "total misuse" of impact factor and similar indicators (Seglen, 1997; Larivière et al., 2016; Zhang et al., 2017). Like it is known that self-citation may distort rankings of low-cited journals, but it almost does not affect top titles (Fassoulaki et al., 2002), so journals with higher IFs probably reflect with more confidence the quality of their single papers measured by citations. In general, the characteristics of the leading journals are uniformity and robustness.

The question raised here is seminal and other hypotheses on inverse correlation between IF and Gini coefficient may be suggested and tested in the future studies.

## Conclusion

Making science involves an element of randomness. To the pool of metaphors of science may be added a concept of "science as a casino". Scientist is a gambler in different ways. Even the initial point of making science, the answer to the question "is my new idea correct?" is a kind of game. Verifying may take much time, attention, financial resources and no one guarantees that the final result will not be *zero*.

Sending a manuscript to a journal, spending your time in discussions with referees is another aspect of the 'science casino', another game. But even after 'accept' decision is received, the gambling is not finished. Actual citations obtained by the paper are also subject to randomness. As Waltman and Traag (2017) suppose, "the skewness of a journal citation distribution <…> does not result from large differences in the values of the articles in a journal. Instead, it results from the inaccuracy of citations as an indicator of the value of an article. As a consequence of this inaccuracy, articles that have a similar value may have very different numbers of citations."

We tried to explore these casinos (journals) and find correlation between their level ('prestige' proxied by IF) and risk an author has if he publishes there. Risk of not receiving 'the average' promised by IF which a scientist sees at the door when he enters the casino.

We found that, in general, higher the stakes, lower the risk of this game. If an author succeeded to pass casino's security, that is referees and editorial policy of a highly cited journal, it is very probable that his paper receives due attention of the readers and, in the long run, gets citations close to average of this leading journal.

The phenomena studied in this paper are based not on simple citation impact measures (this is rather trivial already), but on the "second-order effects", the oscillation of citedness around the average. We use Gini coefficient as a measure of the amplitude of these oscillations. All the results received are similar as for psychology, so for its hard-science and social-science counterparts, neuroscience and sociology.

We have checked intuitively expected relationship between Gini and share of uncited journal's papers and found that the correlation is strong indeed. This means that uncited papers contribute greatly to the unevenness of citation distribution of a whole journal.

The analysis of the correlation between Gini coefficient and IF of a journal revealed some interesting effects. What is more important, these findings have an evident practical application for scientists who plan their publication career.

First of all, in each of the three disciplines covered by our research the journals may be noticed which have almost equal IFs and, at the same time, different Gini indexes. For example, psychology category contains *Journal of Business and Psychology* (IF=2.250, G=0.694) and *European Journal of Work and Organizational Psychology* (IF=2.208, G=0.521). They are highlighted in Figure 2. They are rather close in their scope and sometimes a researcher may think of choosing where to submit his manuscript. Higher Gini means higher risks for a paper. It is a risk in both directions: a paper in such a journal may get much less citations than is "promised" by IF. Generally, there is even a risk of remaining uncited at all. But, on the other hand, in such journals there is a higher probability that a paper may receive significantly more citations than IF's average. This means that, based on his personality characteristics and intentions, a scientist may choose journals with higher or lower risk level when their "average quality" (identified by IF) is the same.

More rationally, we may recommend researchers to consider publishing their *best* papers in journals with high risk level. There they will have a chance to have 'a big win'. The works of moderate quality should rather be sent to a journal with low citation discrimination between papers. There 'no loss' situation is more probable. Once again, we speak here of the journals with *the same* IF.

Next, our finding that in psychology and adjacent disciplines Gini coefficient is inversely related to IF also implies some important information for building an author publication career.

Although somewhat criticized, the approach of publishing in journals with as high IF as possible—until a paper is good enough to overcome the rejection rate barrier—may be justified (cf. Tregoning, 2018). It appeared that titles with higher IF not only demonstrate higher average citedness of their papers, but also increasingly guarantee obtaining this 'average' by individual works published in them.

What is important here, is that it is not just 'a numbers game'. If we believe that citations received by a paper is a valid proxy for its visibility in the scientific world, this means that there are some journals where only a small handful of papers become visible (and maybe even become read) while there are others which almost guarantee some visibility by the mere fact of publishing in them. Of course, this is of practical importance to an author.

To the best of our knowledge, the second-order citation criteria were never applied to authors' publication strategies. The (possible) relation between author's personality and his choice of a journal by bibliometric characteristics has never been addressed and was not studied in the literature. Later it is to be explored if the regularities found in this study also hold for other science domains. Another vital research question for future studies is whether the regularities found here are robust enough over time.